\documentclass[sigconf, nonacm, pdfa]{acmart}

\usepackage[a-2b]{pdfx}
\newcommand\vldbdoi{XX.XX/XXX.XX}
\newcommand\vldbpages{XXX-XXX}
\newcommand\vldbvolume{18}
\newcommand\vldbissue{4}
\newcommand\vldbyear{2024}
\newcommand\vldbauthors{\authors}
\newcommand\vldbtitle{\shorttitle} 
\newcommand\vldbavailabilityurl{https://github.com/yichao-yuan-99/Vortex}
\newcommand\vldbpagestyle{empty} 

\usepackage{subcaption}
\usepackage{tikz}
\usepackage{enumitem}
\usepackage{comment}

\usepackage{multicol}
\usepackage{balance}
\usepackage[absolute]{textpos}
 
\newcommand{\THISWORK}{{\fontfamily{lmss}\selectfont
Vortex}}
\newcommand*\circled[1]{\tikz[baseline=(char.base)]{\node[shape=circle,draw,inner sep=0.85pt, fill=black] (char) {\textcolor{white}{#1}};}}

\begin{document}
\begin{textblock}{16}(3,1)
{\normalsize \normalfont \textit{Authors’ version; to appear in the Proceedings of 51th International Conference on Very Large Databases (VLDB 2025).} }
\end{textblock}
\title{Vortex: Overcoming Memory Capacity Limitations in GPU-Accelerated Large-Scale Data Analytics}

\author{Yichao Yuan}
\affiliation{%
  \institution{University of Michigan}
  \city{Ann Arbor, Michigan}
  \state{USA}
}
\email{yichaoy@umich.edu}

\author{Advait Iyer}
\affiliation{%
  \institution{University of Michigan}
  \city{Ann Arbor, Michigan}
  \state{USA}
}
\email{adviyer@umich.edu}

\author{Lin Ma}
\affiliation{%
  \institution{University of Michigan}
  \city{Ann Arbor, Michigan}
  \state{USA}
}
\email{linmacse@umich.edu}

\author{Nishil Talati}
\affiliation{%
  \institution{University of Michigan}
  \city{Ann Arbor, Michigan}
  \state{USA}
}
\email{talatin@umich.edu}






\begin{abstract}
Despite the high computational throughput of GPUs, limited memory capacity and bandwidth-limited CPU-GPU communication via PCIe links remain significant bottlenecks for accelerating large-scale data analytics workloads.
This paper introduces \THISWORK, a GPU-accelerated framework designed for data analytics workloads that exceed GPU memory capacity. A key aspect of our framework is an optimized IO primitive that leverages all available PCIe links in multi-GPU systems for the IO demand of a single target GPU.
It routes data through other GPUs to such target GPU that handles IO-intensive analytics tasks. This approach is advantageous when other GPUs are occupied with compute-bound workloads, such as popular AI applications that typically underutilize IO resources.
We also introduce a novel programming model that separates GPU kernel development from IO scheduling, reducing programmer burden and enabling GPU code reuse.
Additionally, we present the design of certain important query operators and discuss a late materialization technique based on GPU's zero-copy memory access.
Without caching any data in GPU memory, \THISWORK\ improves the performance of the state-of-the-art GPU baseline, Proteus, by 5.7$\times$ on average and enhances price performance by 2.5$\times$ compared to a CPU-based DuckDB baseline.
\end{abstract}

\maketitle

\pagestyle{\vldbpagestyle}
\begingroup\small\noindent\raggedright\textbf{PVLDB Reference Format:}\\
\vldbauthors. \vldbtitle. PVLDB, \vldbvolume(\vldbissue): \vldbpages, \vldbyear.\\
\href{https://doi.org/\vldbdoi}{doi:\vldbdoi}
\endgroup
\begingroup
\renewcommand\thefootnote{}\footnote{\noindent
This work is licensed under the Creative Commons BY-NC-ND 4.0 International License. Visit \url{https://creativecommons.org/licenses/by-nc-nd/4.0/} to view a copy of this license. For any use beyond those covered by this license, obtain permission by emailing \href{mailto:info@vldb.org}{info@vldb.org}. Copyright is held by the owner/author(s). Publication rights licensed to the VLDB Endowment. \\
\raggedright Proceedings of the VLDB Endowment, Vol. \vldbvolume, No. \vldbissue\ %
ISSN 2150-8097. \\
\href{https://doi.org/\vldbdoi}{doi:\vldbdoi} \\
}\addtocounter{footnote}{-1}\endgroup

\ifdefempty{\vldbavailabilityurl}{}{
\vspace{.3cm}
\begingroup\small\noindent\raggedright\textbf{PVLDB Artifact Availability:}\\
The source code, data, and/or other artifacts have been made available at \url{\vldbavailabilityurl}.
\endgroup
}

\section{Introduction}

GPUs, with their massively parallel architecture, offer high computational power and memory throughput, making them an attractive choice for accelerating large-scale data analytics. 
However, a significant limitation is the memory capacity of GPUs, which is typically constrained to tens or low hundreds of gigabytes in modern hardware. 
In contrast, CPU memory has reached capacities of multiple terabytes. 
Transferring data from CPU memory to GPU memory is often bottlenecked by the limited bandwidth of interconnect links such as PCIe. 
This combination of capacity-limited GPU memory and bandwidth-limited CPU-GPU communication significantly restricts the acceleration potential of GPUs, as real-world data analytics workloads often require processing large datasets that by far exceed the GPU memory capacity.

Prior works addressing this issue can be broadly categorized into two approaches: (1) utilizing multiple GPUs~\cite{heavyai, mg-join-sigmod-2021, multi-gpu-sort-sigmod-2022}, and (2) leveraging CPU-side memory for data processing~\cite{hetexchange-vldb-2019, HERO-VLDB-2017, GDB-TDBSys-2009, mordered-vldb-2022, Ocelot-VLDB-2014, Ocelet-VLDB-2013}. 
The first approach employs multiple GPUs to horizontally scale memory capacity. 
However, this also scales computational resources, maintaining a constant compute-to-memory ratio, and may lead to under-utilization of expensive GPU resources. 
In the second category, some works stream data from CPU memory to GPU~\cite{hetexchange-vldb-2019, HERO-VLDB-2017, GDB-TDBSys-2009}, but still face the bandwidth limitations of PCIe links. 
Additionally, the CPU-GPU hybrid execution approach~\cite{mordered-vldb-2022, Ocelot-VLDB-2014, Ocelet-VLDB-2013} needs to deal with the large disparity in computational power between CPUs and GPUs, risking leaving GPU resources idle when CPU falls on the critical path.
This raises a critical question: \textit{how can we exploit the massively parallel GPU architecture to accelerate data analytics workloads that exceed GPU memory capacity?}

In this paper, we present \THISWORK—a GPU-accelerated framework for large-scale data analytics that addresses memory capacity limitations from a fresh perspective. 
\THISWORK\ has two crucial design goals. 
First, it processes workload sizes that significantly exceed GPU memory capacity. 
\textit{It assumes no data caching in GPU memory before the execution of any query.}
Second, \THISWORK\ is designed to have IO scheduling independent of GPU kernel optimization, alleviating the burden on the GPU programmer.
This unique programming model also maximizes the reuse of optimized GPU code.

The design of \THISWORK\ is structured into three layers. 
With the evolution of multi-GPU systems, we identify a unique opportunity to leverage the IO resources of all GPUs on such systems to transfer data to a single GPU executing data analytics. 
Based on this, at a bottom design layer, we propose an optimized IO primitive that fully exploits the bandwidth available from all PCIe links and inter-GPU communication fabric to transfer data from CPU memory to the GPU at high speeds. 
{\color{black}
This design is motivated by the observation that today's data processing platforms serve both data analytics and AI workloads to support intelligent workflows and diverse needs from users. 
AI workloads tend to be compute-bound, leaving their IO resources underutilized. 
We aim to co-locate compute-bound workloads and data analytics on the same multi-GPU server to process hybrid requests from users.
Our primitive repurposes these idle IO resources from other GPUs to forward data to the target GPU that handles IO-intensive data analytics.


}

We further present the design of an IO-decoupled programming model that clearly separates the development of GPU kernels from IO scheduling at a middle design layer.
In this model, programmers write optimized GPU kernels under the assumption that data is readily available, even for workloads that far exceed GPU memory capacity. 
Our primitive handles IO scheduling and orchestration independently.
This approach significantly simplifies the complex process of GPU code development, promotes the reuse of existing code, and allows for independent exploration of optimization strategies for both kernel execution and IO scheduling. 
Finally, at a top design layer, we implement a set of tailored query operators utilizing our IO primitive and programming model. 
We demonstrate data partitioning and compute orchestration strategies through case studies of sort and hash join, and discuss late materialization based on GPU's zero-copy memory access technique, which further improves query performance.

To demonstrate the effectiveness of \THISWORK, we use AMD's multi-GPU system with 4 GPUs connected to a single CPU socket; however, our techniques are vendor-agnostic. 
We conduct two sets of experiments: (1) a single GPU running data analytics with the other GPUs idle to uncover maximum performance gains, and (2) a single GPU running data analytics while the other three GPUs handle various AI inference tasks to assess the impact on AI workloads when their IO resources are engaged. 
On the end-to-end Star Schema Benchmark~\cite{star-schema-sigmod-02} with a scaling factor of 1000, the first set of experiments shows that \THISWORK\ outperforms the CPU-based DuckDB by 3.4$\times$ and GPU-based Proteus by 5.7$\times$ on average. 
In the second set of experiments, our IO primitive causes only a marginal slowdown of 6.8\% on average for AI workloads. 
Furthermore, we compare the price performance of \THISWORK\ with a CPU-only DuckDB solution that shows benefits ranging from 1.5$\times$--4.2$\times$, depending on the workload.
The key contributions of \THISWORK\ are summarized below.

\begin{itemize}[leftmargin=*]
    \item An optimized IO primitive that utilizes untapped PCIe bandwidth from multiple GPUs for high-speed CPU-GPU data transfer.
    \item An IO-decoupled programming model that separates GPU kernel development from IO scheduling, promoting GPU code reuse.
    \item 
    Tailored query operator implementations with efficient data partitioning and compute orchestration, as well as a late materialization technique based on GPU's zero-copy memory access that further boosts the query performance.
    \item \THISWORK—an end-to-end GPU-accelerated framework that enhances the performance and price-performance of DuckDB by 3.4$\times$ and 2.6$\times$, respectively.
\end{itemize}

\section{Background and Motivation} \label{section:background_motivation}
This section discusses the effort of accelerating database workloads with GPUs systems, challenges faced by existing approaches, and improvement opportunities exploited by our design.

\subsection{Background: Capacity-limited GPU Memory}
Data analytics is a complex and resource-intensive task that involves processing large volumes of data, necessitating substantial computational power. 
The massively parallel architecture of GPUs presents a compelling option for data analytics. 
However, a significant limitation is their on-device memory capacity. 
For instance, next-generation high-end GPUs offer up to 192 GB of memory per card, which is considerably less than the 6 TB maximum DRAM supported by CPUs~\cite{mi300x, genoa}. 
The distinct characteristics of GPU and CPU memory technologies, such as HBM versus DDR, impose physical constraints that make scaling GPU memory capacity challenging. 
Consequently, the disparity between CPU and GPU memory capacities is likely to persist, posing a crucial design question: \textit{How can we utilize the high computational bandwidth of GPUs for data analytics given the limited GPU memory available?}

\subsection{Background: Prior Attempts in Processing Large Datasets using GPUs}


While many prior proposals focus on optimizing for the case where the entire dataset can be stored in GPU memory~\cite{tcudb-sigmod-2022, tqp-vldb-2022, crystal-sigmod-20}, they fall short to effectively handle larger datasets due to GPU memory capacity limit.
This significantly impairs their ability to handle the vast volumes of data typical in real-world applications.
Prior research explores two approaches to accelerate data analytics with GPUs: (1) utilizing multiple GPUs to increase the size of GPU-side memory and (2) using CPU-side memory to hold data.

\noindent
\textbf{Utilizing multiple GPUs.}
High-bandwidth inter-GPU interconnects enable the treatment of memory in multiple GPUs as a unified, large memory pool, thereby allowing for horizontal scaling of GPU memory capacity. 
Systems can take advantage of this fact to hold more data in GPU-side memory~\cite{heavyai, mg-join-sigmod-2021, multi-gpu-sort-sigmod-2022}.
However, this approach is not without drawbacks. 
It necessitates the acceptance of bundled GPU compute resources when the primary issue is memory capacity. 
Such an inflexible strategy may lead to resource under-utilization and consequently increase overall costs.
Additionally, there is a limit to the number of GPU cards a single node can support, typically up to 8, which means the total aggregated memory capacity remains significantly less than what CPUs can provide.

\noindent
\textbf{Using CPU-side memory to hold data.}
This category includes the systems that target CPU-side DRAM to hold large amounts of data. 
Such systems typically adopt a CPU-GPU hybrid solution that utilizes both GPU cores and CPU cores for high performance.
To keep GPU busy, some of them~\cite{hetexchange-vldb-2019, HERO-VLDB-2017, GDB-TDBSys-2009}
choose to stream data from CPU memory to GPU through the CPU-GPU PCIe link.
Identifying the limited bandwidth of PCIe links as a major bottleneck, some other works opt to place part of the data in GPU memory and the rest in CPU memory~\cite{mordered-vldb-2022, Ocelot-VLDB-2014, Ocelet-VLDB-2013}.
The GPU can first process its local data and then handle the work streamed from the PCIe link, which reduces the amount of data that goes through the narrow off-device IO link.
However, we argue that CPU-GPU IO is still a fundamental issue for such solutions.
While the GPU is much faster than the CPU, it has access to less data due to its smaller memory capacity, which creates load imbalance between the CPU and the GPU.
As a result, after the GPU finishes processing its small portion of data, such a solution has to stream data through the PCIe link to keep the GPU running. 
The low IO bandwidth makes it hard to keep up with the speed GPU ingests data.

\begin{figure}
    \centering
    \includegraphics[width=0.43\textwidth, trim={0.25cm 0.0cm 0.25cm 0.05cm}, clip]{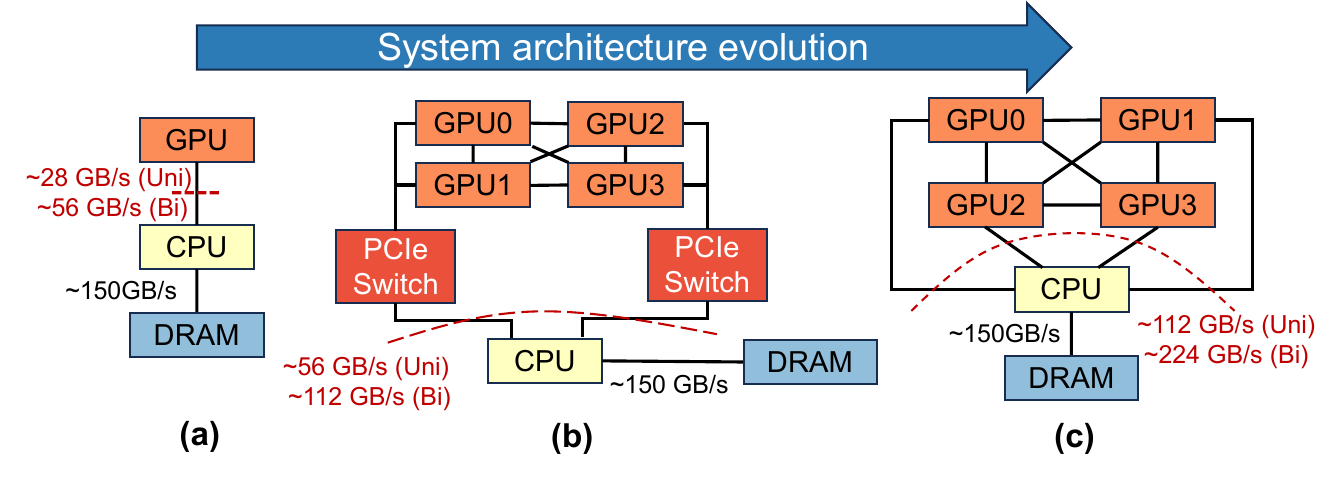}
    \caption{Evolution of GPU system topology.}
    \label{fig:gpu-sys-arch}
\end{figure}

\subsection{Background: Evolution of GPU Systems}
To understand the CPU-GPU data transfer bottleneck and identify new optimization opportunities, it is essential to examine the evolution of GPU system hardware architecture. 
Figure~\ref{fig:gpu-sys-arch}(a) illustrates a classic single GPU system topology, which aligns with conventional understanding. 
In this configuration, the GPU is connected to the CPU via a single PCIe link, while the CPU connects to its memory through DDR channels. 
Considering the common standards of PCIe 4.0 and DDR4-3200, which are prevalent in contemporary systems, the CPU typically has eight memory channels. 
The PCIe link provides approximately 28GB/s bandwidth in one direction and up to 56GB/s in both directions due to its full-duplex nature. 
However, this bandwidth is significantly lower than the 150GB/s bandwidth achievable by the CPU, thereby making CPU-GPU data transfer a critical bottleneck.

To further scale the computing power using multiple GPUs, Figure~\ref{fig:gpu-sys-arch}(b) depicts another system topology found in instances such as AWS p3 and p4~\cite{aws-p3-topo, aws-p4-topo}. 
Due to the limited number of PCIe lanes in older generations of CPUs, supporting four GPUs requires connecting to the CPU through PCIe switches. 
In this configuration, each pair of GPUs shares a PCIe x16 link via these switches, preventing concurrent communication with the CPU at full bandwidth. 
Consequently, the bandwidth between the CPU and GPUs is limited to 56GB/s unidirectionally and 112GB/s bidirectionally. 
While the bidirectional bandwidth approaches the 150GB/s achievable by CPU DRAM, the unidirectional bandwidth remains insufficient by comparison.
In addition to the PCIe links between CPUs and GPUs, modern systems also feature point-to-point high-speed communication between multiple GPUs via technologies such as NVLink (NVIDIA) and Infinity Fabric (AMD).

More recently, server-class CPUs such as AMD's Milan, Rome, and Genoa have begun supporting more PCIe lanes, leading to an architecture where each GPU is directly connected to the CPU through separate PCIe links, as shown in Figure~\ref{fig:gpu-sys-arch}(c).
With all four GPUs connected to the CPU, the aggregated unidirectional PCIe bandwidth is comparable to CPU DRAM bandwidth, and the bidirectional bandwidth even exceeds CPU DRAM bandwidth. 
\textit{Given that CPUs often struggle to fully utilize their DRAM bandwidth, it is time to re-evaluate the assumption that CPU-GPU communication bandwidth is always a bottleneck.}

\subsection{Opportunity: Scaling GPU IO Resources Independently from Compute} \label{subsection:scaling_gpu_io_independently}

Following the topology in Figure~\ref{fig:gpu-sys-arch}(c), a straightforward method to increase CPU-GPU communication bandwidth is to use all GPUs simultaneously. 
However, this approach falls short in resolving the IO bottleneck since it scales both IO and computational power equally. 
The GPU still processes data faster than it can be transferred, maintaining the IO as a bottleneck. 

Our \textit{key observation} is that, unlike data analytics, not all GPU workloads are IO resource-bound. 
For example, modern AI tasks, such as Large Language Model (LLM) inference, are primarily compute-bound and rarely require substantial CPU-GPU data transfer bandwidth. 
Besides, there is a recent trend that data processing platforms not only host classic data analytics services but also AI applications to support users' AI-empowered workflow~\cite{biswal2024text2sqlenoughunifyingai, fan2024surveyragmeetingllms, golatkar2024cprretrievalaugmentedgeneration,xu2023seadendtoendtexttosqlgeneration, hui2021improvingtexttosqlschemadependency} or diverse computation needs that mix AI and data analytics demands from multiple users~\cite{chase-ai, fargo-llm-assitant}.

We aim to exploit the increasingly prevalent multi-GPU systems to more efficiently serve this trend in the future.
We observe a \textbf{\textit{unique opportunity}} to re-distribute the IO resources with multiple GPUs that serve hybrid data analytics and AI workloads. 
Such a system can dedicate the compute of a small number of GPUs to IO-bound data analytics workloads using all GPUs' IO bandwidth. 
It can then use the compute of the remaining GPUs for compute-bound AI workloads while utilizing their IO resources as well.

\textit{In this work, we address the IO bottleneck for GPU-accelerated data analytics by utilizing the compute resources of one GPU and the IO resources of all GPUs in a multi-GPU system.} 
Our approach targets a ``cold-start'' scenario for the GPU - all data resides on CPU memory before the execution starts, therefore the utilization of GPU is not constraint by GPU memory size.
Nevertheless, such an approach may also generalize to more than one GPU for data analytics in other settings, which we leave as future work.

\subsection{Background: Data Transfer Modes on GPU}
\label{sec:SDMA}
GPUs have two primary methods for accessing off-device data: (1) System Direct Memory Access (SDMA) engines and (2) zero-copy memory access.
SDMA is a dedicated hardware component separate from the compute units, responsible for performing data transfers. 
It operates using \texttt{Memcpy*()} API calls in CUDA/HIP to initiate I/O operations, without affecting the ongoing compute tasks on the GPU. 
However, SDMA incurs a fixed overhead for transfer initiation and typically requires a coarse-grained transfers for efficiency, usually on the order of 10s of megabytes.
In contrast, zero-copy memory access allows compute units to directly access CPU-side data at cache line granularity (typically 64/128 bytes). 

\noindent
\textbf{Terminology.}
In the remainder of the paper, the following terminology will be used.
A multi-GPU system is conceptualized as being divided into two sides by the PCIe links.
The side containing the GPUs is referred to as the \textit{Device} side.
The side containing the CPU and DRAM is referred to as the \textit{Host} side.
The direction of data transfer from the Host to the Device is abbreviated as \textit{H2D}.
The direction of data transfer from the Device to the Host is abbreviated as \textit{D2H}.
We use the term \textbf{on-core} to refer to concepts that apply to the case where all data fits GPU memory, and \textbf{out-of-core} for the cases where the dataset is larger than GPU memory.

\section{\THISWORK\ Design Overview} \label{section:design_overview}

We design \THISWORK---a framework that alleviates the IO bottleneck for GPU-accelerated data analytics by tapping into the opportunity discussed in \S\ref{subsection:scaling_gpu_io_independently}.
\THISWORK\ fully engages IO resources on a multi-GPU system. 
Crucially, \THISWORK\ uses a single GPU for computation, and multiple GPUs for data transfers; CPU is only used for orchestrating these data transfers.
The design of \THISWORK\ is divided into three layers as detailed below.

\noindent
\textbf{Optimized IO Primitive (\texttt{Exchange}) (\S\ref{sec:Exchange-IO-primitive})}
Because of non-ideal runtime and hardware behaviors (detailed in \S\ref{sec:io-challenges}), it is challenging to fully leverage the IO bandwidth provided by hardware.
In the presence of these behaviors, we design an efficient IO primitive \textit{\texttt{Exchange}} that allows a target GPU to utilize its neighboring GPUs' SDMA engines and PCIe bandwidth for data transfers.

\noindent
\textbf{IO-Decoupled Programming Model (\S\ref{sec:IO-decoupled-model}). }
Unlike prior works~\cite{triton-join, pump-up-volume, gowan-ipdpsw-2018, sioulas-icde-2019, rui-vldb2020} that integrate the design of IO primitives with compute kernels, we advocate for a fundamentally different approach.
We argue that such an integrated approach is challenging and hinders the reuse of GPU code. 
By leveraging our high-throughput \texttt{Exchange} primitive, we introduce an efficient programming model that decouples IO scheduling from on-GPU kernel design. 
This separation facilitates the reuse of existing code designed only for on-core GPU data processing. It also enables the independent design and exploration of next-generation IO primitives without concerns about the compute kernels.

\noindent
\textbf{Tailored Query Execution (\S\ref{sec:query-execution}).}
With our IO primitive and programming model, we leverage state-of-the-art on-core libraries AMD rocPRIM~\cite{rocprim} and Crystal~\cite{crystal-sigmod-20} to build a set of high-performance out-of-core query operators.
Although we only showcase how to apply our techniques to these libraries, the idea is generally applicable to other execution libraries.

\THISWORK's programming model abstracts away the IO orchestration, but query operators still need to specify their data partitioning strategy for out-of-core processing.
For certain operators, including selection, projection, and join with only one table larger than GPU memory, we can
trivially divide the out-of-core table into multiple chunks and then stream the data onto the GPU.

We illustrate how we implement query operators using the existing on-core GPU libraries with two complex operators that require tailored data partitioning strategy: sort (\S\ref{sec:design-sort}) and hash join with both tables larger than the GPU memory (\S\ref{sec:design-join}).
In \S\ref{sec:design-ssb}, we also introduce a late materialization technique utilizing zero-copy memory accesses as a complementary optimization.

\section{SDMA-based IO Primitive Design}
\label{sec:Exchange-IO-primitive}

\begin{figure}
\centering
\begin{minipage}{.3\linewidth}
    \includegraphics[width=\linewidth]{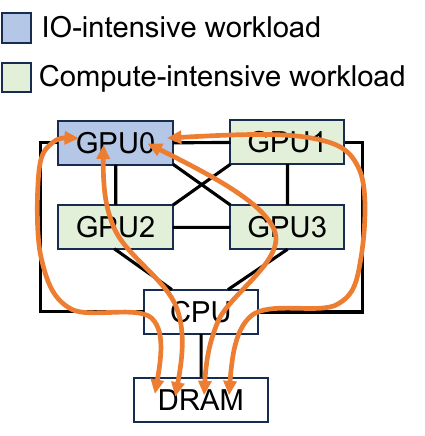}
    \caption{Multiple paths between GPU0 and CPU.}
    \label{fig:io-primitive-flow}
\end{minipage}
\begin{minipage}{.66\linewidth}
    \includegraphics[width=\linewidth]{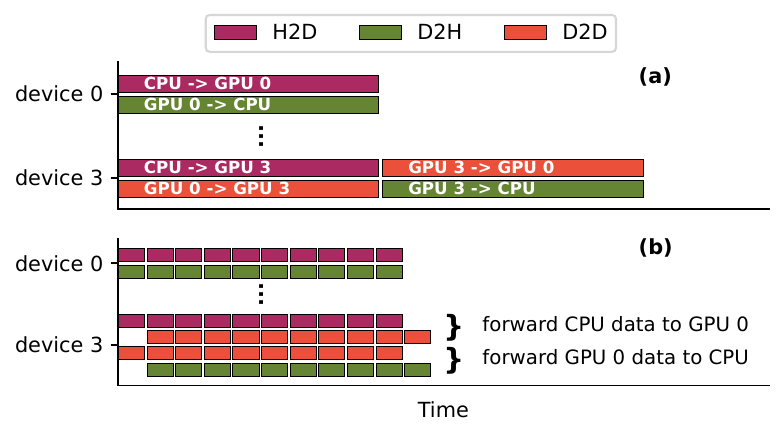}
    \caption{Pipelining data forwarding.}
    \label{fig:io-primitive-pipeline}
\end{minipage}
\end{figure}

\subsection{\textbf{High-level Idea}} \label{sec:io-high-level}
We use an example system with four GPUs to illustrate our high-level idea.
As shown in Figure~\ref{fig:io-primitive-flow}, in multi-GPU systems, data can reach a specific GPU from the CPU-side DRAM through multiple paths. 
In this example, GPU0 handles IO-intensive applications, such as data analytics, while the other three GPUs execute compute-intensive tasks like LLM inference.
GPU0 can leverage the IO bandwidth of its neighboring GPUs through inter-GPU links:
beyond its own PCIe links to the CPU, GPU0 can communicate with the CPU indirectly by using the SDMA engines on the other three GPUs as forwarders, utilizing a small fraction of their GPU memory as buffers. 
These forwarding activities do not significantly affect the normal execution of the other three GPUs, as their SDMA engines and PCIe links are typically underutilized. 
Consequently, this approach significantly enhances the IO capability of GPU0 with minimal impact on the performance of the other GPUs.

Each forwarding GPU needs to first receive data from the source, either GPU0 or the CPU (depending on the direction of data transfer), into a buffer on its GPU memory and then forward the received data to the destination. 
To fully utilize the PCIe bandwidth, the data transmission process needs to be packetized and pipelined.
Figure~\ref{fig:io-primitive-pipeline}(a) illustrates a naive scheduling approach for this process. 
In this approach, the forwarding GPU waits until all data has reached its memory before forwarding it to the destination, failing to take advantage of the full-duplex nature of PCIe links. 
Moreover, a significant amount of memory must be allocated as buffers on the forwarding GPUs, which reduces the capacity available for the compute-intensive kernels.
In contrast, Figure~\ref{fig:io-primitive-pipeline}(b) shows a pipelined schedule that overlaps IO in both directions. 
This approach requires only a small buffer on the forwarding GPU capable of holding a single packet, optimizing bandwidth usage and minimizing memory overhead on the forwarding GPUs.

\subsection{\textbf{Challenges}} \label{sec:io-challenges}
Both NVIDIA's CUDA and AMD's HIP runtimes allow applications to submit a set of concurrent tasks on different \texttt{Streams} and specify dependencies among them through \texttt{Events}~\cite{cuda, hip}.
A straightforward implementation of the idea presented above is to build a Directed Acyclic Graph (DAG) using these two APIs, allowing the runtime to manage data movement in a dataflow order.
In this DAG, nodes represent data packet transfer commands between GPUs launched on different \texttt{Streams}, and edges represent \texttt{Events} that enforce the forwarding order.
However, in practice, we find two challenges to achieve desired performance as listed below.

\noindent
\textbf{Challenge \#1. Head-of-line blocking.} 
Although operations are submitted to different streams, they must traverse multiple layers of software queues before ultimately being enqueued into a limited set of hardware-managed queues~\cite{olmedo-rtas-2020, otterness-rtns-2021}. 
The execution engines process operations from these queues in FIFO order. 
This FIFO order can lead to head-of-line blocking between operations on different streams. 
As a result, part of the Host-to-Device IO operations are unintentionally blocked by the runtime in our baseline solution.

\noindent 
\textbf{Challenge \#2. Non-uniform IO bandwidth.}
When both directions of all PCIe links are used simultaneously for data transfer, the combined IO bandwidth requirement exceeds the CPU-side memory controller's capacity, resulting in some PCIe links not achieving their maximum bandwidth. 
This leads to non-uniform bandwidth availability across different paths, and the bandwidth achieved by each link is highly dependent on the activity on other links.
For example, when all GPUs perform H2D data transfer and GPU0, 1, and 2 transfer data in D2H, the achievable H2D bandwidth on GPU0 and 2 is only around 8GB/s, and around 20GB/s on GPU1.
This is significantly less than the 28GB/s bandwidth when there is no D2H traffic.
In our test machine with 4 AMD MI100s, data traffic in the D2H direction consistently outperforms H2D traffic when competing for bandwidth, making H2D links more susceptible to interference from other data transfers.
This issue causes load imbalance in data transfers. 
When transferring data in both directions using our baseline solution, all H2D paths are assigned the same number of equally-sized packets, yet they complete in significantly different times, leading to underutilized PCIe links at the end. 

\noindent
\textit{These challenges motivate the necessity of a more sophisticated solution to fully utilize the potential of PCIe links.}

\subsection{\textbf{Proposed Design}}
We follow two design principles to address the challenges above.

\begin{enumerate}[leftmargin=*]
    \item \textit{Only submit requests that will be immediately executed by the GPU.}
    By ensuring that, no task waits inside all queue levels, and we can eliminate unpredictable head-of-line blocking. 
    \item \textit{Consider all concurrent data transfers and implement comprehensive flow control.} 
    Given the irregular nature of IO bandwidth, a holistic flow control mechanism that accounts for traffic in both directions is essential to maintaining overall load balance.
\end{enumerate}

\begin{figure}[t]
    \centering
    \includegraphics[width=\linewidth]{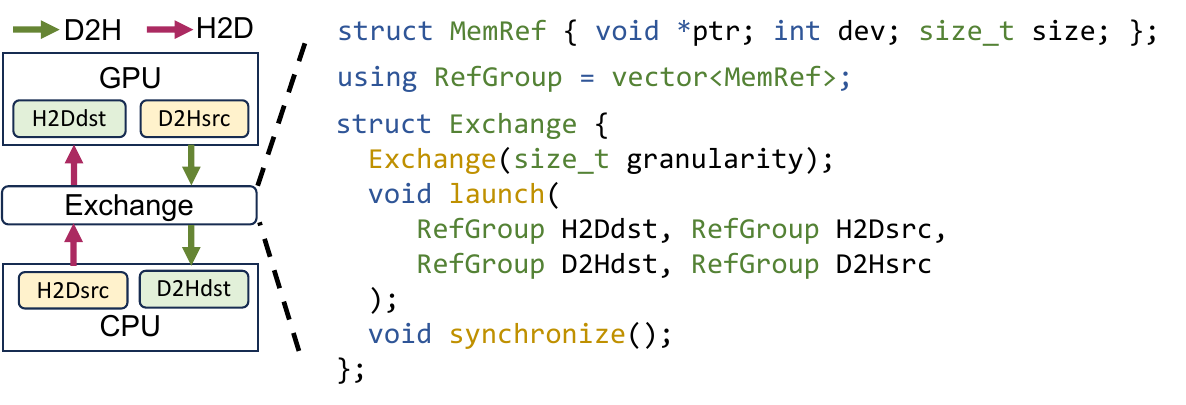}
    \caption{The interface of \texttt{Exchange} operation.}
    \label{fig:io-primitive-interface}
\end{figure}

\noindent
\textbf{Interface.}
We introduce a key primitive, \texttt{Exchange}, based on the observations discussed, as illustrated in Figure~\ref{fig:io-primitive-interface}. 
The \texttt{Exchange} operation asynchronously transfers data between the CPU and the target GPU while simultaneously managing data movement in both directions. 
This transfer is pipelined at a specified packet size granularity. 
By scheduling H2D and D2H data movements together, \texttt{Exchange} enables the underlying scheduler to manage both directions and perform flow control effectively.
Unlike traditional \texttt{Memcpy} APIs, which require source and 
destination memory to be contiguous, \texttt{Exchange} only mandates that the sizes of the source and destination be identical. 
For example, it allows copying 4 chunks of 2 GB data from CPU DRAM to an 8 GB region in GPU memory.
The design of \texttt{Exchange} elevates CPU-GPU data movement from basic runtime APIs to a more sophisticated library component. 
This design philosophy mirrors practices in databases, which maintain their own buffer pools rather than relying on lower-level \texttt{mmap}~\cite{crotty2022you}, and aligns with the use of collective communication primitives like NVIDIA’s NCCL and AMD’s RCCL.

\begin{figure}[t]
    \centering
    \includegraphics[width=0.88\linewidth]{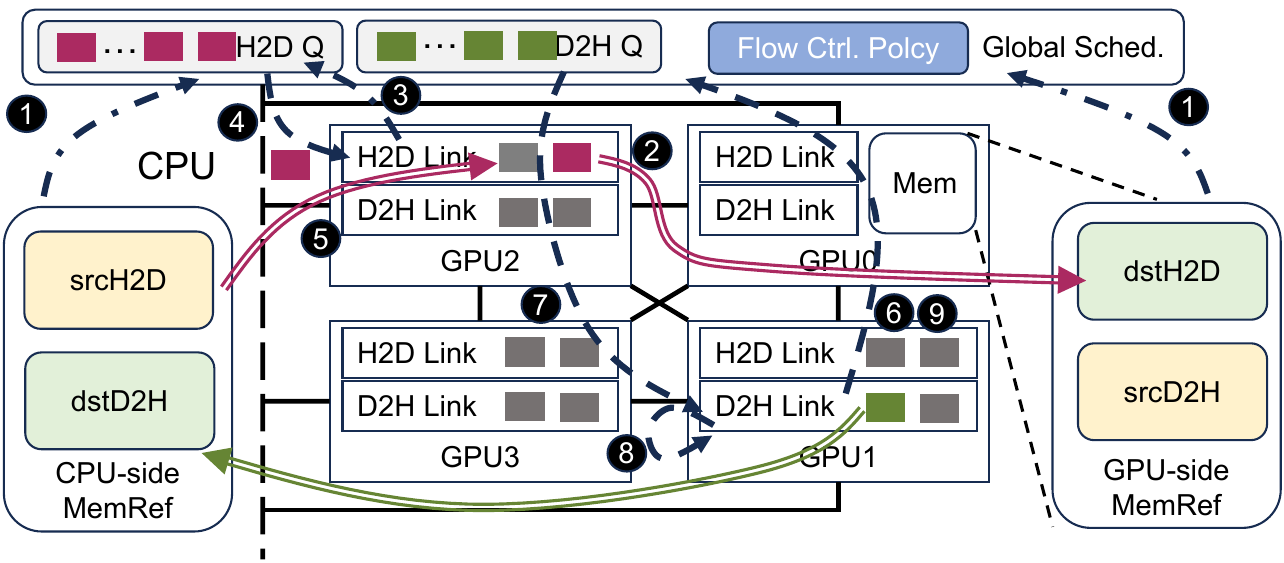}
    \caption{The implementation details of \texttt{Exchange} operation.}
    \label{fig:exchange-details}
\end{figure}

\noindent
\textbf{Implementation details.}
Our implementation of the \texttt{Exchange} operation includes both link-level workers for data packet transmission and a global scheduler to ensure balanced load distribution among links.
Figure~\ref{fig:exchange-details} provides an example with GPU0 as the target GPU. 
Each GPU initializes two link workers dedicated to H2D and D2H data transfers, respectively.
The link worker on forwarding GPUs, referred to as the \textit{indirect link}, maintains two buffers sized to the packet granularity and alternates between them to enable pipelined packet forwarding.
In contrast, the link worker on the target GPU, known as the \textit{direct link}, functions as a proxy to the underlying runtime memory copy API.

The global scheduler manages two task queues for H2D and D2H packets, populated from the arguments to \texttt{Exchange}. 
Flow control is enforced by a policy that dictates whether a link can dequeue a task from its respective queue. 
Given that H2D traffic consistently suffers from bandwidth competition on our test machine, our policy ensures that the D2H queue does not drain faster than the H2D queue; otherwise, the links are restricted from popping tasks from the D2H queue.
A more universal policy could involve maintaining a balanced number of tasks in each queue.
Below, we provide a detailed walk-through of this process, illustrated in Figure~\ref{fig:exchange-details}.

When \texttt{Exchange} is instantiated, it initializes a pair of links for each GPU and a global scheduler. 
Upon invoking the \texttt{launch()} method, the global scheduler partitions the \texttt{dstH2D} and \texttt{srcH2D} data at the packet size granularity and populates the H2D queue (\circled{1}). 
Similarly, it populates the D2H queue.
All links are then activated and continuously pop tasks from their respective queues. 
Each link follows a work cycle until the queue is empty. 
In each cycle, the indirect link pushes any buffered packets from the previous cycle to the destination (\circled{2}) and attempts to dequeue a task from the queue (\circled{3}). 
If permitted by the flow control policy, the link retrieves a task (\circled{4}) and then transfers a packet from the source to its buffer (\circled{5}).
If the flow control policy rejects the pop request (\circled{6}), the link receives a stall signal (\circled{7}), waits for a short period (\circled{8}), and retries the task pop (\circled{9}). 
The direct link operates similarly but moves data directly to the destination. 
Empirically, we set the waiting period to 10us, which is a few percent of the packet transfer time.
Each link is blocked at the end of each iteration until all IO operations issued by the link are completed. 
This ensures that at the beginning of the next iteration, the execution units used by the link are idle, allowing any requests to be executed immediately rather than waiting in the queue, adhering to the design principle of immediate execution (1).

\section{IO-Decoupled Programming Model}
\label{sec:IO-decoupled-model}

\subsection{\textbf{High-level Idea}}
With CPU-GPU data transfer bandwidth as a primary bottleneck, many prior works design a tightly coupled CPU-GPU IO schedule with compute kernels~\cite{triton-join, pump-up-volume, gowan-ipdpsw-2018, sioulas-icde-2019, rui-vldb2020}.
While these approaches have achieved promising results, they complicate the design space by intertwining IO management with GPU kernel design, often leading to ad-hoc optimization techniques. 
Furthermore, this strategy limits the reuse of optimized and validated kernels provided by GPU vendors, which are not designed for out-of-core processing.

We propose that, given \texttt{Exchange}'s significant mitigation of the IO bottleneck, adopting a decoupled programming model that separates IO management from on-GPU computation is advantageous. 
This model allows for a clear separation of concerns, facilitating code reuse while maintaining high performance.
In this approach, programmers can develop \textit{data-parallel}~\cite{data-parallel} operations by using regular on-GPU kernels and specifying how to partition the data to fit into GPU memory, leaving the framework managing all IO orchestration.
Programmers can develop GPU kernels with a traditional single-GPU model in mind, reusing code initially designed for GPU-memory-fitted data. 
\THISWORK\ is designed for data analytics, thus cannot accelerate algorithms/data structures involving data access jumps exceeding the GPU memory size (i.e., tens of GBs).
However, in the scope of data analytics, many operations are naturally data-parallel, like projection and aggregation, and others can be decomposed into steps of data-parallel sub-operations, as we will demonstrate for sort and join (\S\ref{sec:design-sort} and \S\ref{sec:design-join}).

\begin{figure}[t]
    \centering
    \includegraphics[width=0.95\linewidth]{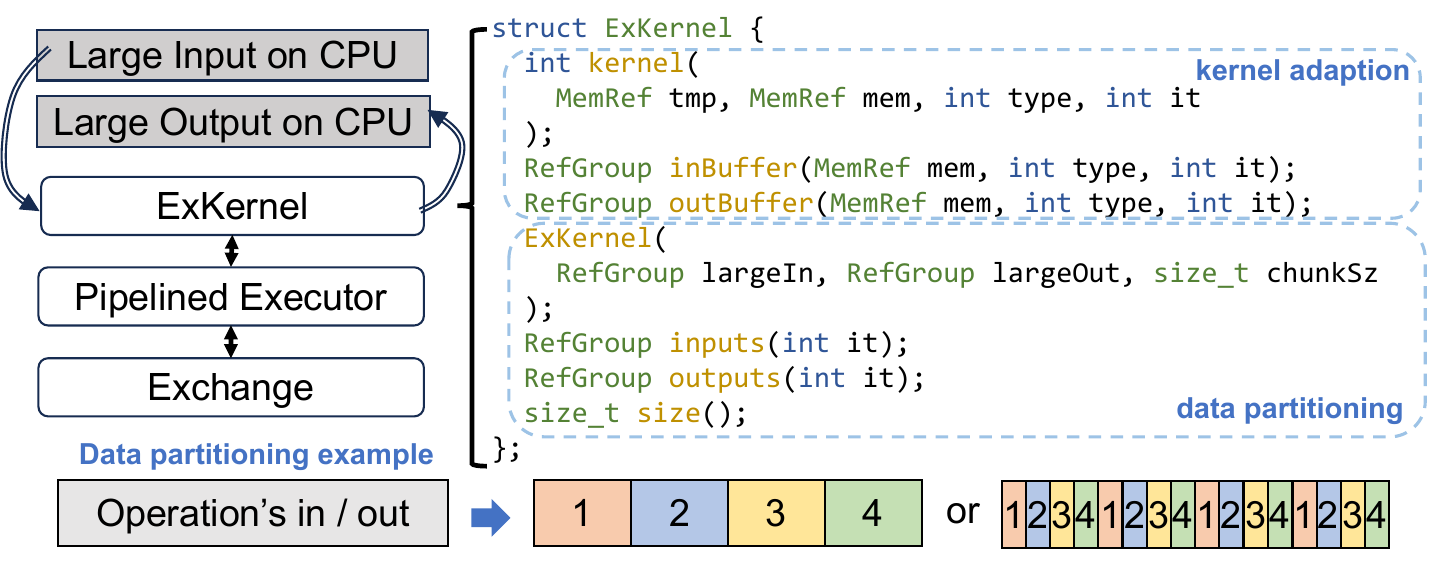}
    \caption{IO-decoupled Programming Model at a High Level, the interface for \texttt{ExKernel}, and examples of how \texttt{ExKernel} may partition its input or output. }
    \label{fig:pipeline-high-level}
\end{figure}

The right-hand side of Figure~\ref{fig:pipeline-high-level} illustrates the proposed programming model. 
In this model, programmers extend their existing on-GPU kernels into \texttt{ExKernel}s, short for \texttt{\textbf{Ex}}tended \texttt{\textbf{Kernel}}.
\texttt{ExKernel}s handle large datasets that exceed GPU memory by processing data stored in DRAM and then saving the results back to DRAM. 
These \texttt{ExKernel}s are executed by a \texttt{Pipelined Executor}, which leverages \texttt{Exchange} to manage off-GPU IO. 
This approach allows \texttt{ExKernel}s to efficiently process out-of-core data while fully utilizing PCIe bandwidth.

\subsection{\textbf{Design Details of \texttt{ExKernel}}}
The interface of \texttt{ExKernel} is depicted in Figure~\ref{fig:pipeline-high-level}.
An \texttt{ExKernel} comprises two sets of methods.
(1) \textit{Data Mapping}: this set of methods directs the executor on how to process the large dataset in chunks on the GPU using the on-GPU kernel.
(2) \textit{Kernel Adaptation}: this set of methods defines how the on-GPU kernel interacts with the executor, specifying how to receive input and produce output.

\noindent
\textbf{Data Mapping.}
Efficient processing of datasets larger than GPU memory requires chunk-wise handling, where each chunk fits into GPU memory. 
The \texttt{ExKernel} \textit{maps} inputs and outputs into \textit{chunks} of size \texttt{chunkSz}, ensuring each chunk fits within the allocated memory. 
The chunks are accessible through the \texttt{inputs()} and \texttt{outputs()} methods, while \texttt{size()} indicates the total number of chunks to process. 
During this process, data in DRAM remains stationary; only a mapping table is created to associate data with its respective chunk. 
A chunk can be either a contiguous memory region or a collection of regions. 
Contiguous chunks are suitable for embarrassingly parallel operations, such as element-wise addition. 
In contrast, non-contiguous chunks are used when processing outputs from other \texttt{ExKernel}s, as demonstrated in our sort (\S\ref{sec:design-sort}) and hash join (\S\ref{sec:design-join}) designs.

\noindent \textbf{Kernel adaption.}
The \texttt{kernel()} method wraps the on-GPU kernels designed solely for processing data within GPU memory. 
It provides these kernels with
(a) \texttt{mem}: a memory region containing both input and output data,
(b) \texttt{type}: a code indicating the layout of \texttt{mem},
(c) \texttt{it}: an index specifying the chunk of data currently being processed, and
(d) \texttt{tmp}: a memory area for temporary storage during execution.
The layout of \texttt{mem} is defined by the \texttt{inBuffer()} and \texttt{outBuffer()} methods. 
These methods, given the \texttt{type} code and index \texttt{it}, return specific regions of \texttt{mem} for input and output.
Upon completion, \texttt{kernel()} returns a \texttt{type} code that can be used with \texttt{outBuffer()} to locate the output and with \texttt{inBuffer()} to access the next input.

\noindent
\textbf{Output management.}
\texttt{outputs()} can cover a smaller or larger range of memory than \texttt{inputs()} for operations with different input/output sizes.
\texttt{kernel()} may populate \texttt{outputs()} during execution for operations with indefinite output size. 
However, programmers need to be conservative on \texttt{inputs()} chunk size to avoid overflowing GPU memory buffers.
More details are in our technical report.
Developing a more versatile input/output management mechanism for indefinite-sized outputs is left as future work.

\begin{figure}[t]
    \centering
    \includegraphics[width=0.75\linewidth]{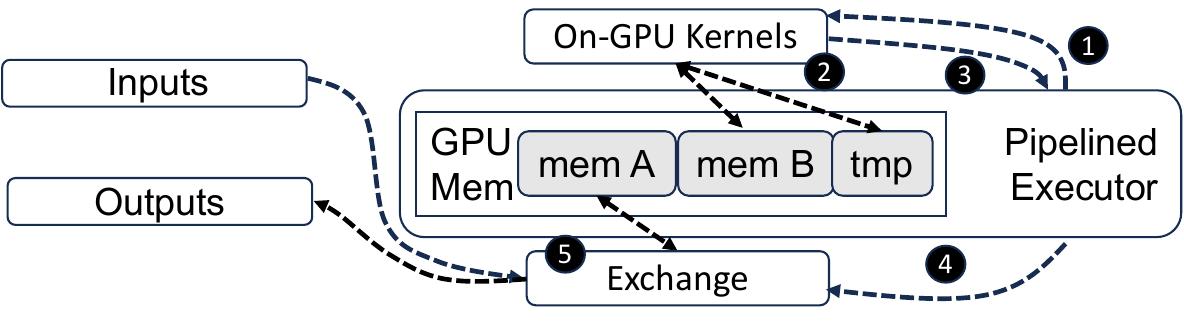}
    \caption{Implementation details of \texttt{ExKernel} execution.}
    \label{fig:pipeline-details}
\end{figure}

\subsection{\textbf{Design Details of \texttt{Pipelined Executor}}}
\label{sec:executor}
We illustrate the execution of an \texttt{ExKernel} using the pipeline executor with an example in Figure~\ref{fig:pipeline-details}.
The executor divides GPU memory into three sections: two memory buffers, \texttt{mem A} and \texttt{mem B}, each allocated 16 GB from the 32 GB available on the AMD MI100, and a \texttt{tmp} memory for temporary storage.
The computation is performed in a software-pipelined manner~\cite{softpipe-pldi-1988}, running through multiple cycles. 
During the $n$th pipeline cycle $C_n$:
(1) The executor processes the $(n-1)$th chunk, which was previously loaded into \texttt{mem B}, by invoking \texttt{kernel(it=n-1)} (\circled{1}).
(2) The kernel reads from and writes to \texttt{mem B} and \texttt{tmp}, leaving the output in \texttt{mem B} (\circled{2}).
(3) The executor records the \texttt{type} code returned by \texttt{kernel()} to determine the input and output buffers for the next cycle (\circled{3}).
(4) Concurrently, an \texttt{Exchange} operation is performed over \texttt{mem A}, with source and destination determined by the recorded \texttt{type} code from the previous cycle (\circled{4}).
(5) The input for the $n$th chunk is loaded from the CPU, and the output of the $(n - 2)$th chunk is saved to the CPU (\circled{5}).

\section{Tailored Query Execution}
\label{sec:query-execution}
In this section, we present the details of sort and hash join, as well as late materialization based on zero-copy memory access.

\begin{figure}[t]
    \centering
    \includegraphics[width=0.75\linewidth]{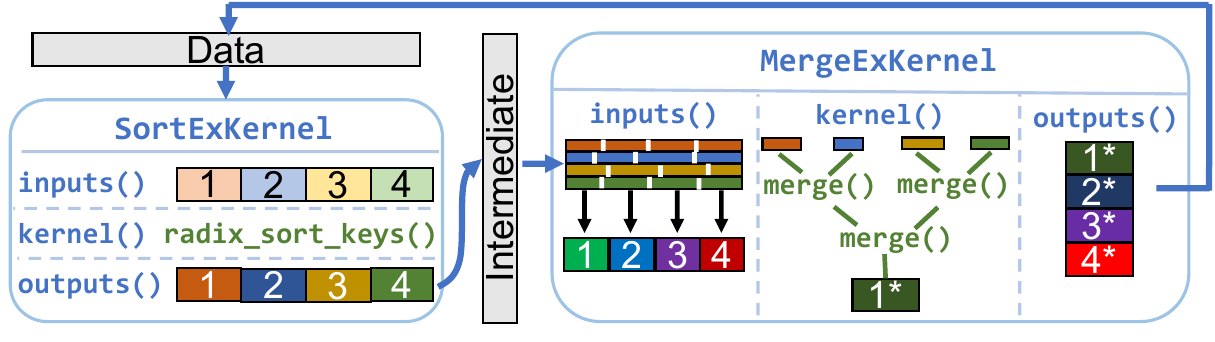}
    \caption{The implementation details of sort operation upon the IO-decoupled programming model.}
    \label{fig:sort-flow}
\end{figure}

\subsection{Case Study -- Sort}
\label{sec:design-sort}
In this section, we illustrate the design of the sort operator in the context of sorting a large array of 64-bit integers that exceeds GPU memory capacity, which adapts the classic external sort algorithm~\cite{external-sort}.
Our approach, as shown in Figure~\ref{fig:sort-flow}, employs two \texttt{ExKernel}s. 
The \texttt{SortExKernel} is responsible for partitioning the input array and sorting each chunk individually. 
The sorted chunks are then stored in an intermediate buffer on the CPU. 
Subsequently, the \texttt{MergeExKernel} merges these sorted chunks and writes the final output back to the original data location. 
To highlight the code reuse capabilities of our programming model, we utilize existing GPU kernels provided by vendors, avoiding the need to develop custom GPU kernels.

\subsubsection{\textbf{Integration with Vendor-maintained GPU Primitives.}}
Both AMD and NVIDIA offer highly optimized parallel primitives for their GPUs through libraries like rocPRIM~\cite{rocprim} and CUB~\cite{cub}, respectively. 
These libraries, extensively tested and continuously updated with cutting-edge techniques~\cite{onesweep-2022, mergepath-2012}, ensure optimal performance. 
In our work, we utilize rocPRIM primitives due to our focus on AMD GPUs, noting that its interface closely mirrors that of NVIDIA's CUB.

We briefly outline the rocPRIM primitives used and their input/output conventions.
\textit{Double Buffer} consists of two memory regions, designated as \texttt{current} and \texttt{alternate}. 
It is a standard construct in GPU sorting operations.
\texttt{radix\_sort\_key()} operates on a double buffer where \texttt{current} contains the input array. 
After sorting, \texttt{current} holds the sorted array, and the roles of \texttt{current} and \texttt{alternate} are swapped.
\texttt{radix\_sort\_pair()} sorts pairs of keys and values using two double buffers. 
It outputs the sorted keys and values into the \texttt{current} regions of each buffer.
\texttt{merge()} merges two sorted arrays into a third output array, taking pointers to the input and output arrays as arguments.
These primitives facilitate efficient data processing by leveraging optimized GPU operations and memory management.

\subsubsection{\textbf{Design Details of \texttt{SortExKernel}}}
\label{sec:SortExOperation}
The goal is to sort the given array in multiple chunks, which are subsequently merged.

\noindent
\textbf{Data Mapping.}
The input is divided into several contiguous chunks that fit within the on-GPU buffer, as illustrated in Figure~\ref{fig:sort-flow}. 
The output on the CPU side is mapped similarly, with each chunk containing a sorted sub-array.

\noindent
\textbf{Kernel Adaption.}
We wrap \texttt{sort\_by\_key()} to handle the sorting operation on the GPU.
To integrate the double buffer data structure with the executor, the buffer's marker is used as the \texttt{type} code. 
The methods \texttt{inBuffer()} and \texttt{outBuffer()} interpret the provided \texttt{mem} as a double buffer, with the \texttt{current} buffer indicated by the \texttt{type} code. 
They return this \texttt{current} buffer accordingly. 
Similarly, the \texttt{kernel()} method reconstructs the double buffer based on the \texttt{type} code and invokes \texttt{sort\_by\_key()}. 
After \texttt{sort\_by\_key()} completes the sorting and updates the buffer marker, \texttt{kernel()} returns the new marker as the updated \texttt{type} code.

\subsubsection{\textbf{Design Details of \texttt{MergeExKernel}}}
This \texttt{ExKernel} merges multiple sorted chunks produced by the \texttt{SortExKernel} into a single, fully sorted array in a single pass.

\noindent
\textbf{Data Mapping.}
The input to this operation consists of \( N \) sorted chunks, each with a size of \( C \). 
The objective is to regroup these chunks into \( N \) new chunks \( P_i \) (\( 0 \leq i < N \)) such that \(\text{max}(P_i) \leq \text{min}(P_{i + 1})\) for all \(0 \leq i < N - 1\). 
Each new chunk will contain \( N \) sorted segments, which can then be efficiently merged on the GPU.
We devise an efficient algorithm based on binary search, the details of which can be found in our technical report~\cite{vortex-technical-report}.
In Figure~\ref{fig:sort-flow}, the white points on \texttt{MergeExKernels}'s \texttt{inputs()} segment the input chunks and reorganize them into the desired partitions.
The output is then partitioned to store the final sorted result.

\noindent
\textbf{Kernel Adaption.}
The on-GPU kernels are tasked with merging the sorted segments within each partition. 
This process is performed in a tree-like order, as depicted in Figure~\ref{fig:sort-flow}, using the \texttt{merge()} kernel from rocPRIM.
We employ the double buffer layout described in the sorting step. 
During each iteration, the kernel merges pairs of segments from the \texttt{current} buffer and writes the results to the \texttt{alternate} buffer. 
After completing the merge operations in an iteration, the buffers are swapped. 
The management of the \texttt{type} code, \texttt{inBuffers}, and \texttt{outBuffers} follows the same approach as detailed in \S\ref{sec:SortExOperation}.

\begin{figure}[t]
    \centering
    \includegraphics[width=0.9\linewidth]{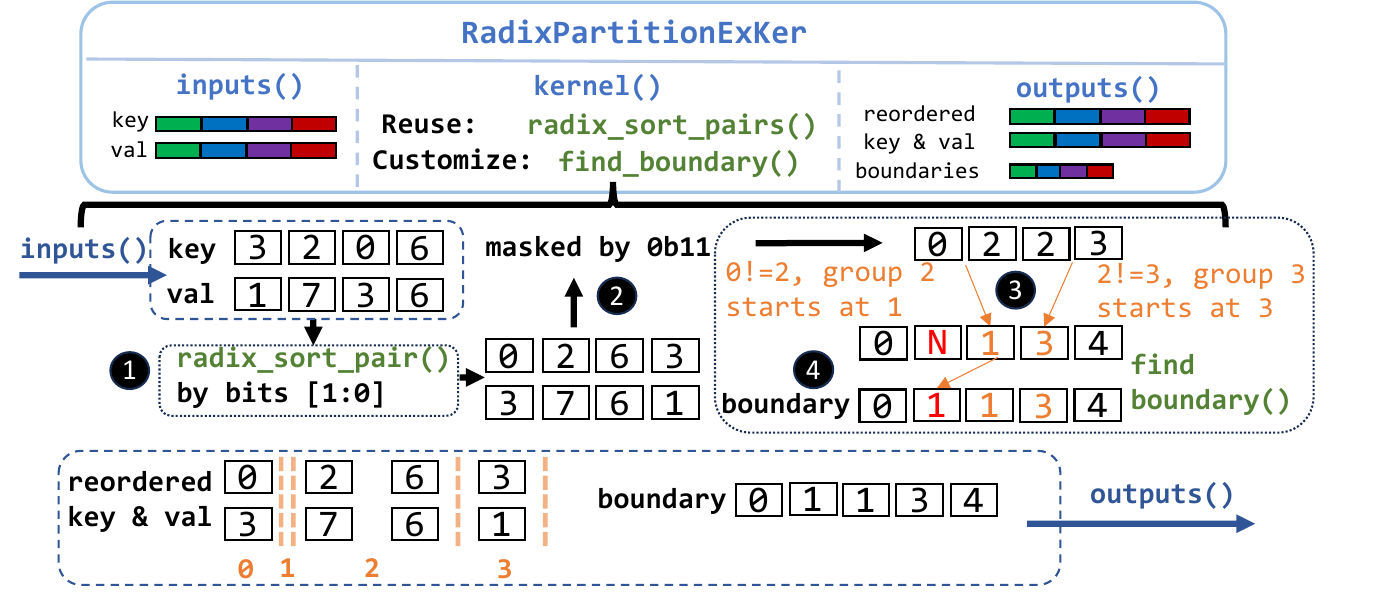}
    \caption{Implementation details of \texttt{RadixPartitionExKer}. }
    \label{fig:radix-partition-flow}
\end{figure}

\subsection{Case Study -- Hash Join}
\label{sec:design-join}
Next, we discuss the design of the hash join operator on an equality predicate where both tables exceed the GPU memory.
We refer to the two tables as  \( A \) and \( B \) and the join predicate as \texttt{A.key == B.key}.
Based on the classic idea of radix-partitioned hash join~\cite{partitioned-join-vldb99}, Triton join~\cite{triton-join} proposes a solution that surpasses CPU performance, relying heavily on a specialized high-bandwidth CPU-GPU NVLink configuration. 
In contrast, we achieve even better results using our \texttt{Exchange} primitive and an IO-decoupled programming model, which utilizes standard PCIe links to access CPU-side memory.
For this case study, we follow the problem setup outlined in~\cite{triton-join}. 
Both tables \( A \) and \( B \) have equal sizes, with each row consisting of an 8-byte unsigned integer tuple \texttt{<key, val>}. 
The \texttt{key} in table \( B \) contains foreign key referencing \texttt{A}, with the keys distributed uniformly.
Following the idea of radix-partitioned hash join, our solution includes a \texttt{RadixPartitionExKer} to cluster tuples with identical hashes into \textit{groups} and a \texttt{HashJoinExKer} to join the \textit{groups} from both tables with the same hashes together.

\subsubsection{\textbf{Design Details of \texttt{RadixPartitionExKer}}}
Figure~\ref{fig:radix-partition-flow} illustrates the details of the \texttt{RadixPartitionExKer}. 
This \texttt{ExKernel} maps the table into chunks and clusters tuples within each chunk using a hash function. 
Following~\cite{triton-join}, it utilizes the 0 to 23 bits of the key as the hash value, partitioning the table into approximately 16.8 million groups.
While a better hash function can be used to handle the skewed key distribution, this is out of the scope of our paper as we focus on IO optimization.
Tuples in each group share a distinct hash value ranging from 0 to $2^{24} - 1$. 
Groups with the same hash value can be joined independently in the subsequent phase.

\noindent
\textbf{Kernel Adaption.}
We build the on-GPU clustering kernel by integrating the \texttt{radix\_sort\_pair()} primitive from rocPRIM with our custom \texttt{find\_boundary()} kernel, rather than developing it from scratch.
The clustering kernel organizes tuples based on their hash values and generates a \texttt{boundary} array to demarcate group boundaries.
As illustrated in the lower half of Figure~\ref{fig:radix-partition-flow}, consider an example with 3-bit unsigned integers where we use the lower two bits of the keys as hashes to cluster them into 4 groups. 

(1) The executor loads a partition of 4 tuples from the CPU. The kernel uses \texttt{radix\_sort\_pair()} to sort these tuples by the lower two bits of their keys (\circled{1}).
Although the tuples are now clustered, the exact boundaries between groups remain undetermined.
(2) The \texttt{find\_boundary()} kernel generates a \texttt{boundary} array with $N + 1$ elements for $N$ groups, whose $i$th and $(i+1)$th elements tell the boundary of the group with a hash $i$.
Initially, this array contains placeholder values (\textit{e.g.,} \texttt{N} for "Not Available"). 
The kernel masks the keys with \texttt{0'b11} to obtain their hash values (\circled{2}). 
It then performs parallel checks to identify transitions between different hashes.
(3) If a hash $h$ at position $i$ differs from its previous hash, this signifies the start of a new group. 
The position $i$ is recorded in the corresponding $h$th entry of the \texttt{boundary} array (\circled{3}). 
In the example, the second entry remains 1 because the group with hash 1 is empty.
(4) To complete the \texttt{boundary} array, we replace any remaining \texttt{N} entries with the subsequent valid boundary positions, effectively denoting empty groups (\circled{4}). 
With the tuples now clustered and the \texttt{boundary} array populated, the results are stored back to the CPU for further processing.

\noindent
\textbf{Data Mapping.}
The table's \texttt{key} and \texttt{val} columns are mapped into chunks.
When the table is mapped into \texttt{N} chunks, the output will include \texttt{N} chunks of clustered \texttt{key} and \texttt{val}, as well as \texttt{N} chunks of \texttt{boundary} to indicate mark the boundaries for each chunk.

\subsubsection{\textbf{Design Details of \texttt{HashJoinExKer}}}
After both tables are radix partitioned, the join phase processes groups with identical hash values from each table.

\noindent
\textbf{Data Mapping.}
The goal of this step is to ensure that all groups with the same hash value are processed by the on-GPU kernel together. 
We carefully partition the data by binary search over `boundary' chunks, and present more details in technical report~\cite{vortex-technical-report}.

\noindent
\textbf{Kernel Adaption.}
We fully customize the hash join kernel to leverage GPU shared memory for enhanced performance. 
In the partitioning phase, the data is divided into 16.8 million groups that can be processed independently. 
Even with a dataset of 16 billion rows, each group contains approximately 1000 tuples and occupies around 16KB, which fits comfortably within the 64KB shared memory available per GPU core.
As this paper focuses on out-of-core GPU processing, we leave the details of this on-GPU kernel for our technical report~\cite{vortex-technical-report}.

\subsection{Case Study -- Late Materialization}
\label{sec:design-ssb}
Late materialization is a common optimization for column-oriented databases, where only columns referenced by selection/join predicates and tuples not filtered out by selection/join predicates are fully materialized from disk to memory~\cite{abadi2006materialization}. 
\THISWORK~utilizes this strategy to reduce the amount of data transferred from CPU memory to GPU memory. Consider a selection operator based on two predicates $\sigma_a$ and $\sigma_b$ for a table with two columns $a$ and $b$.
If $\sigma_a$ is highly selective, the execution engine only needs to transfer column $a$ from CPU memory to GPU to apply $\sigma_a$ first. Then, it only transfers the values of column $b$ from the tuples that satisfy $\sigma_a$ to GPU to further evaluate $\sigma_b$. 
The join predicates are optimized similarly.


However, such a technique requires fine-grained CPU-GPU data access based on the predicate of each tuple, which is not achievable efficiently by SDMA-based IO primitives (\S\ref{sec:SDMA}).
In contrast, although zero-copy memory access can operate at cache line granularity, it cannot enjoy the idle IO resource on other GPUs due to its need for compute units.
We propose to late materialize a column $c$ with zero-copy access in an \texttt{ExKernel} based on a selectivity estimator $\hat{S_c}$ for that column, which is determined before that \texttt{Exkernel} happens.
If $\hat{S_c}$ is less than an architectural-dependent threshold $TH$, the execution engine will not load column $c$ but let the on-core kernel use zero-copy memory access to retrieve data on demand. 
Otherwise, it will still load $c$ through SDMA-based \texttt{Exchange}.
The threshold $TH$ is determined by GPU LLC cache line size, $C_{l2}$, the size of the accessed element, $E$, as well as the number of GPUs \texttt{Exchange} uses, $N_{exchange}$.
$$
TH = \frac{E}{C_{l2} \times N_{exchange}}
$$
If the selectivity is higher than $TH$, the benefit of selective data access is less than plainly using neighboring GPUs' IO resources.
On our test machine with 4 MI100 GPUs, whose LLC cache line size is 64 bytes, $TH = \frac{4}{64 \times 4} = \frac{1}{64}$, when we are accessing 4-byte integers.
We validate our formula with the following micro-benchmark.
\begin{figure}[t]
    \centering
    \includegraphics[width=\linewidth]{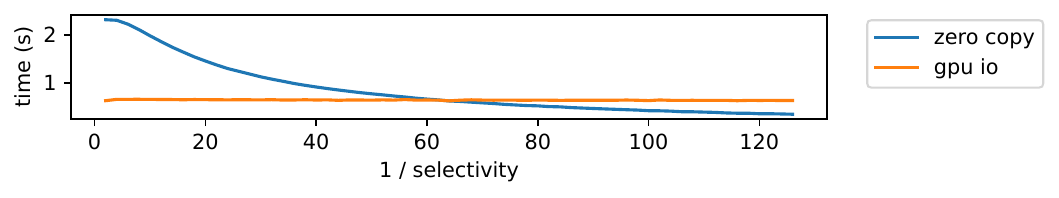}
    \caption{Zero copy vs GPU IO.}
    \label{fig:selectivity-perf}
\end{figure}

\begin{verbatim}
for i in range(16e9)
  sum += pred[i % 2e9] % SEL == 0 ? v[i] : 0
\end{verbatim}
The \texttt{pred} array resides in GPU memory, and \texttt{SEL} is a hyperparameter that is inversely related to selectivity. 
We implement this micro-benchmark using both \texttt{Exchange} and zero-copy data transfer techniques, varying \texttt{SEL} from 1 to 128. 
The results are presented in Figure~\ref{fig:selectivity-perf}. 
Notably, when \texttt{SEL} \(> 64\), zero-copy becomes more efficient. 
This aligns with the threshold \(TH = \frac{1}{64}\), validating the results derived from our formula.

\subsection{Discussion}
\THISWORK\ can be used to support data analytics engines with different processing models.
In this work, we bind \THISWORK\ with Crystal~\cite{crystal-sigmod-20} that uses kernel fusion for query execution, but executing an operator at a time is also viable by chaining multiple \THISWORK\ extended GPU operators. 
For example, libraries like cuDF~\cite{cudf} implement their operators
based on on-GPU compute primitives in Thrust~\cite{Thrust} and CUB~\cite{cub}, which shares similar APIs as the rocPRIM~\cite{rocprim} we use. 
One may take a similar approach described in \S\ref{sec:design-sort} and \S\ref{sec:design-join} to extend the operators with \THISWORK, and chain them for query execution.
The CPU memory can hold the intermediate results.

\section{Evaluation Methodology} \label{section:methodology}

We evaluate \THISWORK\ in two distinct scenarios.
We first run \THISWORK\ while keeping all forwarding GPUs idle to understand the upper bound on the improvement from our IO primitive.
We require that all input/output data be retrieved from and stored in CPU's DRAM, with no GPU-side caching, which represents the most challenging setup for GPU execution. 
Then, we concurrently run \THISWORK\ on the target GPU and deep learning workloads on the forwarding GPUs to analyze the interference introduced by our techniques.
While an alternative to partition system resources is by running data analytics on 25\% of four GPUs, we experimentally show in the technical report~\cite{vortex-technical-report} that this is an inferior design choice.

\subsection{Interference-Free Analysis}
\noindent
\textbf{\texttt{Exchange}.}
We evaluate the efficiency of our \texttt{Exchange} IO primitive by varying both the total volume of data transferred and the granularity of the data packets. 
For benchmarking, we compare it against an in-house implementation that relies solely on the GPU runtime.
This baseline is susceptible to the challenges outlined in \S\ref{sec:io-challenges}.

\noindent
\textbf{Sort.}
In this experiment, we sort 8 billion 8-byte integers and use TBB and PARADIS~\cite{paradis-vldb-2015}, which are state-of-the-art CPU sort implementations from industry and academia, as our baselines. 
Due to the absence of open-source out-of-core GPU sort libraries, we compare \THISWORK\ against in-house GPU baseline. 
The only distinction between this and \THISWORK\ is that the former utilizes the IO resources of a single GPU, whereas \THISWORK\ leverages multiple GPUs.

\noindent
\textbf{Hash Join.}
Following the experimental setup outlined in \cite{triton-join}, we focus on the query below
\begin{verbatim}
SELECT SUM(A.val + B.val) FROM A, B WHERE A.key == B.key;
\end{verbatim}
Because we lack access to machines equipped with CPU-GPU NV-Links, we use the results from \cite{triton-join} as our GPU baseline. 
For our CPU baselines, we employ DuckDB and the CPU hash join implementation from \cite{triton-join}, representing state-of-the-art solutions from both industry and academia.

\noindent
\textbf{Star Schema Benchmark (SSB).}
SSB is an OLAP benchmark widely used by many prior database research~\cite{ydb-2013, crystal-sigmod-20, mordered-vldb-2022, kaibo-vldb-2014, hetexchange-vldb-2019}, composed of 13 queries grouped in 4 query flights that process over a fact table and a few small dimension tables.
These queries commonly filter the dimension table, join them with the fact table, and aggregate the joined results. 

We integrate \THISWORK\ with the on-core query processing library Crystal~\cite{crystal-sigmod-20} to implement queries in SSB.
As the dimension table is small, we load the dimension tables to GPU memory, filter and build the hash table for them, and keep them inside GPU memory.
Then, we partition the fact table and process these partitions one by one by Crystal.
We utilize the late materialization optimization when transferring the fact table to the GPU.
Because the selectivity of each dimension table is known after their hash tables are built, we can use the selectivity of the dimension table as an accurate estimator for the selectivity of each column in the fact table.
Thus, we can optimize the star schema query by late materializing some columns in the fact table during query execution, as long as their selectivity estimator is less than $TH$.

{\color{blue}

}


We evaluate SSB at a scale factor (SF) of 1000, comprising 6 billion rows in the fact table. 
The entire database spans approximately 600 GB, with 144 GB designated as hot data for query processing. 
For benchmarking, we use DuckDB as our CPU baseline and Proteus~\cite{hetexchange-vldb-2019} as our GPU baseline, which is a leading out-of-core GPU database. 
We compare \THISWORK\ against Proteus in two configurations: CPU-GPU hybrid execution mode, and pure-GPU execution mode using default settings.
\THISWORK\ executes each query independently without caching data between queries.

\subsection{Interference Analysis}
As we are borrowing idle IO resources from neighbors that run compute-intensive applications like Deep Learning (DL), it is important to understand the degree of interference caused by our technique in such a multi-tenant system.
We run a range of deep learning workloads from Hugging Face on the forwarding GPUs to assess the impact of \THISWORK\ on system performance.

\noindent
\textbf{Text-to-image Diffusion Model Inference.} 
We use \textit{stable-diffusion-3-medium} (SD3)~\cite{stable-diffusion}, a widely recognized text-to-image generative model that creates images from user prompts. 
Since the diffusion model operates iteratively, we measure performance based on the number of iterations completed per second.

\noindent
\textbf{Text Embedding Generation.}
Text embeddings convert text into high-dimensional vectors and are crucial for text mining and Retrieval-Augmented Generation (RAG) applications, which enhance LLM.
In our evaluation, we use the state-of-the-art text embedding model \textit{e5-mistral-7b-instruct}~\cite{wang2022text, wang2024improving}. 
This model processes text sequences to generate corresponding embeddings. 
We measure its performance by the number of embeddings produced per second.

\begin{figure*}[t]
\centerline{\includegraphics[width=0.89\linewidth]{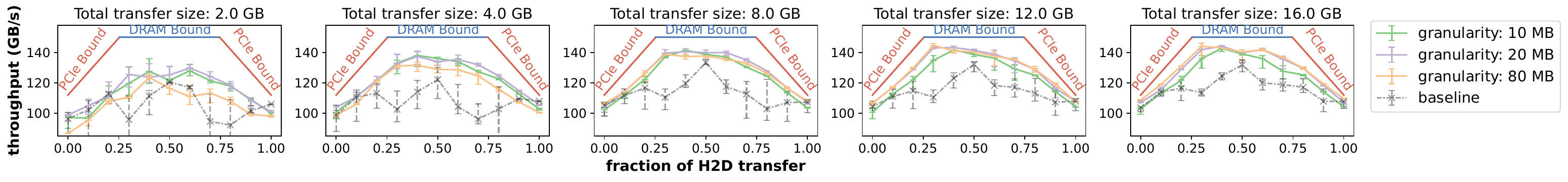}}
\caption{Data transfer throughput achieved by the IO-primitives with different transfer granularity.}
\label{fig:io-bandwidth}
\end{figure*}

\begin{figure*}[t]
    \centering
    \includegraphics[width=0.85\linewidth]{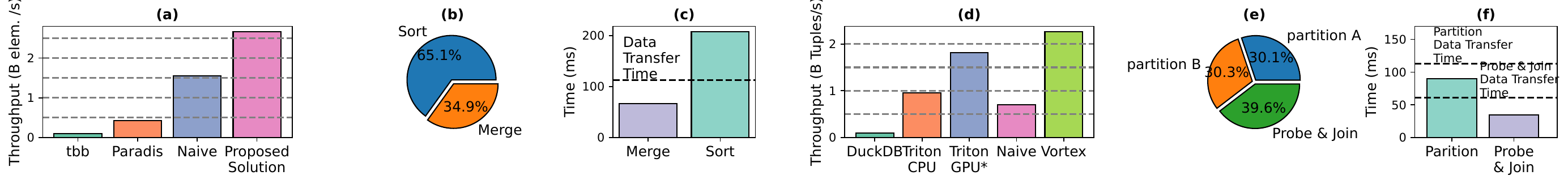}
    \caption{Results for Sort (a-c) and Join (d-e). (a,d) the throughput achieved by different solutions, (b,e) the time breakdown for the \THISWORK\ sort/join, and (c,f) the time taken by on-GPU kernel execution of a typical pipeline stage.}
    \label{fig:sort-perf}
    
\end{figure*}

\noindent
\textbf{LLM Serving.}
Our evaluation of LLM serving focuses on \textit{Meta-Llama-3-8B-Instruct}~\cite{llama3modelcard}. 
LLM inference comprises two distinct phases: the compute-intensive prefill phase, which processes prompts, and the memory-intensive decode phase, which generates text. 
We assess these stages separately, given that they have different workload characteristics and modern LLM serving systems often process them on decoupled GPU pools. 


\noindent
\textbf{Workload Characteristics}
The DL workloads we analyze exhibit varying demands on GPU memory subsystems and compute units. 
Diffusion models, text embedding generation, and the prefill stage of LLM serving are characterized by high arithmetic intensity and are therefore classified as compute-bound applications~\cite{golden2024generativeaillmsimplications, zhao2024prepackingsimplemethodfast}. 
Conversely, the decode stage of LLM serving is memory-bound when processing small batch sizes but shifts to compute-bound as batch sizes increase~\cite{zhang2024flattenquantbreakinginferencecomputebound}. 
A detailed analysis on how these characteristics affect interference is present in \S\ref{sec:interference}.

\subsection{Hardware Configuration}
All the GPU workloads, except Proteus, are run on a dual-socket server with 8 AMD MI100 GPUs connected to two AMD 7V13 processors in groups of 4.
We only use 4 GPUs and one socket for this work.
The GPUs are directly connected to the CPU through PCIe 4.0 links, which provide ~28GB/s maximum bandwidth in each direction.
AMD 7V13 CPU has 8 memory channels running at 3200Mhz, and we measure the maximum DRAM bandwidth at 150GB/s using the STREAM benchmark~\cite{stream-benchmark}.
Because Proteus is developed based on Nvidia GPUs, we run it on Nvidia A40, a GPU comparable to AMD MI100.
We run CPU baselines on a dual-socket server with two Intel Xeon Platinum 8380 processors, which also have 8 memory channels per socket and deliver 150GB/s memory bandwidth.
We choose this server for CPU baselines as they achieve better performance compared to the AMD server.

\section{Evaluation Results} \label{section:restuls}

\subsection{Interference-Free Analysis}
\noindent
\textbf{Performance of the \texttt{Exchange} primitive.}
Figure~\ref{fig:io-bandwidth} illustrates a comparison of the IO throughput achieved by our optimized \texttt{Exchange} and the baseline solution, which solely relies on the GPU runtime. 
We vary the total amount of data transferred from 2GB to 16GB and adjust the packet size from 10MB to 80MB. 
The combination of data size and packet size determines the total number of packets, which in turn affects the number of pipeline stages required for data transfer. 
Too few pipeline stages can lead to significant overhead in the prologue and epilogue phases of the pipeline. 
Conversely, utilizing excessively small packets is also inefficient, as each memory copy incurs a fixed overhead from the runtime, regardless of the transferred data volume. 
Therefore, small packet sizes exacerbate this overhead, making it disproportionately large.

Our solution achieves up to 140GB/s throughput when transferring 8GB or more of data. 
When the total amount of data transferred is small, we observe a decrease in throughput due to the reduced number of pipeline stages. 
As previously explained, this issue cannot be alleviated by simply reducing the packet size. 
For instance, while a packet size of 10MB provides better performance compared to an 80MB packet size when transferring 2GB of data, it delivers lower throughput when the data size exceeds 8GB. 
Empirically, we find that a packet size of 20MB strikes a balance, achieving desirable performance for small and large data transfers.
Consequently, we use a packet size of 20MB for all the applications evaluated below.

Compared to the baseline, which fully relies on the GPU runtime, our solution is not only more performant but also more stable. 
Such a baseline fails to fully exploit the full-duplex capabilities of PCIe links, achieving only about 110-130GB/s throughput when transferring data bidirectionally. 
Additionally, its performance is highly unstable due to the irregular PCIe bandwidth, especially when the CPU DRAM bandwidth becomes saturated.

\noindent
\textbf{Performance of Sort.}
We compare our sort implementation with CPU and GPU baselines in Figure~\ref{fig:sort-perf}(a). 
Our implementation achieves a throughput of 2.7 billion elements per second, which is 27.9$\times$ faster than TBB, 6.3$\times$ faster than PARADIS, and 1.7$\times$ faster than the configuration using only one GPU's IO resources. 
Figure~\ref{fig:sort-perf}(b) provides a breakdown of the sort operation, revealing that 65.1\% of the time is consumed by the \texttt{SortExKernel}. 
This occurs because, after enhancing the IO throughput, the sorting operation becomes bounded by the GPU processing throughput, as illustrated in Figure~\ref{fig:sort-perf}(c). 
While it takes the GPU approximately 208ms to sort a partition of 500 million 8-byte integers, transferring that partition to the GPU using four GPUs' IO resources requires only about 113ms. 
This limitation explains why we do not achieve nearly a 4$\times$ speedup compared to the single GPU IO solution. 
Conversely, the \texttt{MergeExKernel} remains IO-bound, with the on-GPU kernel completing in approximately 67ms.

\begin{figure*}[t]
\centerline{\includegraphics[width=\linewidth]{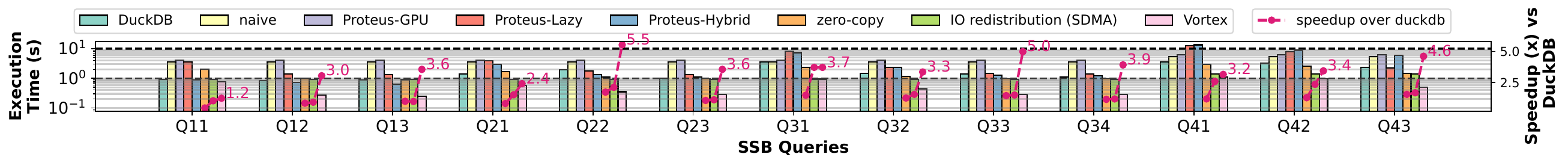}}
\caption{Star Schema Benchmark execution time and speedup.}
\label{fig:ssb-perf}
\end{figure*}
\begin{figure}
    \centering
    \includegraphics[width=0.86\linewidth]{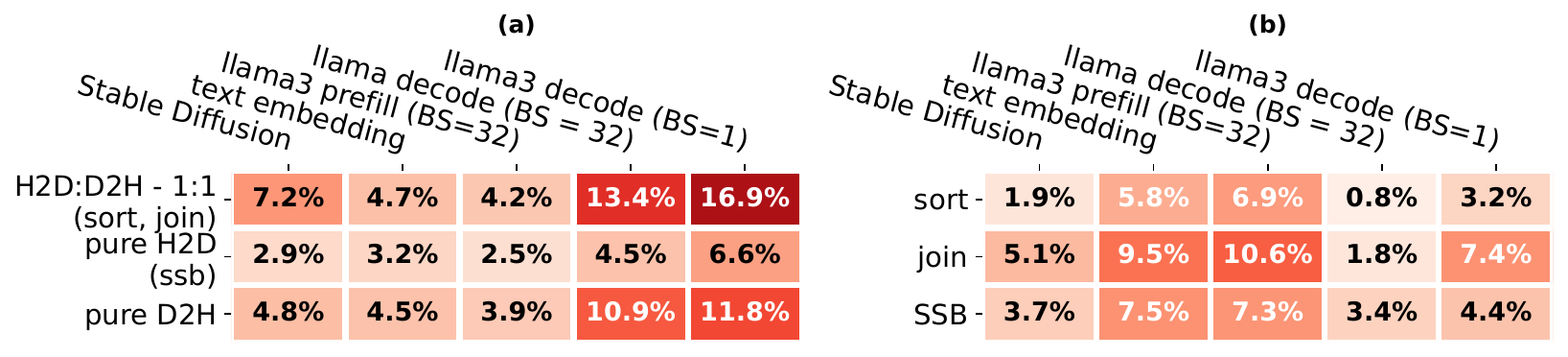}
    \caption{Interference between \THISWORK\ on the target GPU and the deep learning applications on the forwarding GPUs. 
    }
    \label{fig:interference}
\end{figure}
\noindent
\textbf{Performance of Hash Join.}
In contrast to sorting, hash join remains an IO-bound kernel even with our IO optimization technique. 
As shown in Figure~\ref{fig:sort-perf}(d), our solution achieves a throughput of 2.3 billion tuples per second. 
This is 24.1$\times$ faster than DuckDB, 2.4$\times$ faster than Triton Join (CPU), 1.3$\times$ faster than the CPU-GPU-NVLink-based Triton Join (GPU), and 3.2$\times$ faster than the single GPU solution using a standard PCIe link.
The speedup over the single GPU IO solution is more pronounced because all phases of hash join are IO-bound. 
This is evident in Figure~\ref{fig:sort-perf}(f). 
The \texttt{HashJoinExKer} requires only 34ms to complete the on-GPU join kernel, which is significantly less than the 61ms required for data transfer.
Similarly, it takes 90ms to partition a chunk of data, which is transferred in around 113ms. 
All phases scale uniformly with the improvement of IO throughput, as depicted in the time breakdown in Figure~\ref{fig:sort-perf}(e), where they consume a comparable amount of time. 
Notably, \THISWORK\ outperforms Triton Join without using proprietary CPU-GPU interconnects by exploiting untapped PCIe bandwidth.

\noindent
\textbf{Performance of SSB queries.}
Figure~\ref{fig:ssb-perf} illustrates the comparison of SSB query performance between our solution and the baseline approaches.
On average, our solution achieves a 3.4$\times$ speedup over DuckDB, with all data dynamically fetched from CPU DRAM.
When examining individual query flights, the speedup is 2.4$\times$ for Q1.*, 3.6$\times$ for Q2.*, 3.9$\times$ for Q3.*, and 3.7$\times$ for Q4.*. 
The higher speedup observed in Q2.*, Q3.*, and Q4.* is attributed to their inclusion of more complex multi-way joins.
The more complex multi-way join demands higher memory throughput for hash table probing, thus favoring GPU-based solutions more as they can operate in high-bandwidth GPU memory.
The CPU-based solution has to use the limited DRAM bandwidth on hash table probing and fact table reading, while our solution only uses DRAM bandwidth for the latter.
Lightweight queries like Q11 only filter the fact table based on some predicates, whose only DRAM traffic is reading the fact table once.
Thus, the benefit of high-bandwidth GPU memory is minimized, and we observe less speedup.

By comparing the bars of \texttt{navie} and \texttt{Proteus-GPU} with \texttt{DuckDB}, it becomes evident that GPU-based solutions struggle to achieve performance comparable to the CPU-based DuckDB without utilizing our IO optimization technique. 
However, this technique alone is insufficient, as indicated by the comparison between the \THISWORK\ and \texttt{DuckDB} bars. 
It only achieves a 1.6$\times$ speedup against \texttt{DuckDB} because it transfers unused data to the GPU without considering column selectivity. 
While zero-copy can exploit selectivity, it falls short of maximizing throughput because it relies on a single PCIe link. 
Notably, using zero-copy alone results in worse performance than \THISWORK\ .
Our final solution dynamically switches between SDMA-based data transfer for columns with selectivity greater than a threshold \(TH\) and zero-copy data transfer for columns with selectivity less than \(TH\).
Our solution also achieves 5.7$\times$ speedup over \texttt{Proteus-Hybrid}, despite that it uses both CPU and GPU.
It is difficult for such a hybrid solution to divide work between CPU and GPU and efficiently utilize the CPU DRAM bandwidth.
Our solution achieves 6.2$\times$ speedup over \texttt{Proteus-Lazy}, which enhances \texttt{Proteus-GPU} with late materialization techniques.
After we resolve the IO bottleneck and fully utilize CPU-side DRAM, a pure GPU-based solution can achieve highly competitive results.

\subsection{Interference Analysis}
\label{sec:interference}
\noindent
While \THISWORK\ utilizes additional GPUs and their IO resources to forward data to a target GPU, running AI workloads on these auxiliary GPUs can lead to a slowdown of these workloads.
Figure~\ref{fig:interference}(a) presents the slowdown for the AI applications (x-axis) when the IO traffic (y-axis) runs in the background, and (b) shows the slowdown for the \THISWORK\ applications (y-axis) when the deep learning applications (x-axis) run in the background.
(1) Compared to single-direction IO traffic, bidirectional IO traffic has a more significant impact on the performance of foreground applications. This is likely due to the increased stress placed on the memory subsystems of the forwarding GPUs.
(2) Memory-intensive workloads are more susceptible to interference from data forwarding activities, as their performance is constrained by the memory bandwidth available on the GPUs. 
Background data forwarding consumes a portion of the memory bandwidth, leading to an average slowdown of 6.8\%.
Compared to SD3, text embedding generation, and LLM prefilling, LLM decoding experiences a greater degree of slowdown.

Figure~\ref{fig:interference} illustrates that current hardware may not optimize for our IO optimization techniques due to two key observations.
First, although the memory subsystem is theoretically stressed to the same degree in both scenarios, forwarding IO traffic from the device to the host results in a more significant slowdown compared to traffic from the host to the device.
Second, to support the 140GB/s IO throughput we achieved, each GPU incurs an additional memory bandwidth cost of $\frac{140 \times 2}{4} = 70$GB/s, which constitutes only $\frac{70}{1200} \approx 5.8\%$ of the MI100's total bandwidth.
However, empirical observations reveal slowdowns of 7.2\%, 13.4\%, and 16.9\% for \texttt{SD3}, \texttt{Llama3} decoding with a batch size of 32, and \texttt{Llama3} decoding with a batch size of 1, respectively.
We hypothesize that this discrepancy arises because our programming model generates atypical memory traffic that hinders the GPU memory controller's ability to fully utilize bandwidth for the foreground application.

We analyze the slowdown of data analytics applications caused by DL applications on forwarding GPUs. 
As shown in Figure~\ref{fig:interference}, the target GPU experiences less slowdown, with a maximum of 10.4\%. 
However, the slowdown patterns are more irregular compared to forwarding GPUs. 
Text embedding generation and \texttt{Llama3} prefilling cause more interference than \texttt{SD3}, despite all being compute-bound workloads. 
Interestingly, the memory-bound \texttt{Llama3} decoding shows less interference on the target GPU, contrasting with the significant interference on the forwarding GPUs.

\noindent
\textbf{Overall system efficiency.}
Given that our technique can accelerate heavily IO-bound applications by 3 to 4 times, we argue that the system is still more efficient even with a slowdown of up to 16.9\% on the other GPUs.
The improvement of overall system efficiency in a 4-GPU system can be quantified as shown below.
$$
\text{speedup}_\text{sys} = \frac{\text{speedup}_\text{t} * \text{slowdown}_\text{t} + 3 * \text{slowdown}_\text{f}}{4}
$$
The subscripts `t' and `f' denote the target GPU and forwarding GPUs, respectively. 
Consider the scenario where \texttt{SD3} and hash join, both with primarily bidirectional IO traffic, are collocated.
The overall system speedup is $\frac{3.2 * (1 - 0.051) + 3 * (1 - 0.072)}{4} \approx 1.45$.
In our setup, the least favorable combination is running \texttt{Llama3} decoding without batching alongside sort. 
Despite this, we still achieve a modest speedup of$\frac{1.7 * (1 - 0.032) + 3 * (1 - 0.169)}{4} \approx 1.03$ speedup.
Note that these speedup values refer to the entire 4-GPU system. 
For a single GPU, they correspond to speedups of 2.8$\times$ and 1.12$\times$, respectively.

\section{Price Performance Analysis} \label{section:cost_analysis}
This section evaluates the price performance of our proposed system compared to a baseline CPU-only solution.
A key assumption we take is that all input and output data is stored on the CPU-side DRAM without caching on the GPU.
The price of \THISWORK\ is 
$$
C_{\text{\THISWORK}} = C_\text{raw} + C_{\text{raw}} \times \text{tax} = C_\text{raw} +  C_{\text{raw}} \times (\text{tax}_{\text{t}} + 3 \times \text{tax}_{\text{f}}).
$$
$C_{\text{raw}}$ is the raw price of a GPU.
The term $\text{tax}$ is introduced to account for the tax paid for enhanced IO throughput.
This factor takes the interference on the target GPU, $\text{tax}_{\text{t}}$, and the forwarding GPUs,  $\text{tax}_{\text{f}}$, into account, which are calculated as $ \textit{tax}_{\textit{t | f}} = 1 - \frac{1}{\text{slowdown}_\text{t | f}} $.
The price performance of GPU-accelerated \THISWORK\ over a CPU baseline is calculated as $ \textit{price performance} = \frac{C_\text{cpu}}{C_\text{\THISWORK} + C_\text{cpu}} \times \text{speedup} $.
To calculate the price of renting a GPU, we approximate the price with NVIDIA's A100 GPU (as AMD MI100 GPUs are not available from the cloud providers).
Arguably, A100 is a more powerful GPU with a higher price, making our price estimation \textit{conservative}.
The price of renting an A100 GPU is estimated using AWS \texttt{p4d.24xlarge} and \texttt{r5dn.metal} instances.
\texttt{p4d.24xlarge} offers 8 A100 GPUs with a more powerful CPU and larger DRAM capacity than \texttt{r5dn.metal} with no GPU.
Therefore, the upper-bound price of a single A100 GPU is estimated using
$$
C_{\texttt{raw-A100}} = \frac{C_{\text{p4d.24xlarge}} - C_{\text{r5dn.metal}}}{8} = \frac{32.773 - 8.016}{8} \approx 3.1 (\$/\text{h})
$$
Because the CPU on \texttt{r5dn.metal} is weaker than the CPU we use, we use the $C_\text{cpu} = C_\text{r5dn.metal}$ to calculate lower bounds of the baseline price, making our estimation even more conservative.
When the background is running \texttt{llama3} decode without batching that introduces the highest degree of interference, the price performance of \THISWORK\ compared to a CPU for sort, hash join, and SSB is 3.9$\times$, 1.5$\times$, and 2.3$\times$, respectively.
When the background is running \texttt{SD3}, the price performance of \THISWORK\ for sort, hash join, and SSB is 4.2$\times$, 1.6$\times$, and 2.4$\times$, respectively.
On average, \THISWORK\ achieves 2.5$\times$ better price performance over the CPU-based DuckDB.
Notably, the discussed price estimations are conservative. In practice, we expect \THISWORK\ to achieve even better price performance.

\section{Related Work} \label{section:related_works}

\textbf{GPU native query processing systems.}
Many GPU-accelerated systems~\cite{heavyai, mg-join-sigmod-2021, multi-gpu-sort-sigmod-2022} optimize for the case that the entire dataset can be stored in GPU memory, and use CPU-based solutions or stream data to GPU only as fallback plans. 
Such systems focus on improving the performance of on-device GPU kernels while addressing the memory capacity problem by using multiple GPUs.
In contrast, our work enables the GPU to process large datasets in CPU-side memory by borrowing the IO bandwidth from the other GPUs.
\THISWORK\ also directly benefits from techniques that improve on-device GPU kernels~\cite{Funke-sigmod18, crystal-sigmod-20}, as it can reuse existing GPU code through its IO-decoupled programming model.


\noindent
\textbf{CPU-GPU hybrid query processing systems.}
To handle large datasets, multiple systems~\cite{hetexchange-vldb-2019, HERO-VLDB-2017, GDB-TDBSys-2009, mordered-vldb-2022, Ocelot-VLDB-2014, Ocelet-VLDB-2013, GPUQP, FlinkCL} place part or all of the data in CPU memory and use both CPU and GPU to process the query. 
These works include multiple optimizations to enhance operator placement and data placement and reduce the amount of data to be processed.
\THISWORK\ focuses on IO optimization and is orthogonal to prior works in this direction.
\THISWORK's discussion on late materialization emphasizes trading off zero-copy access and our SDMA-based IO primitives and provides a verified analytical formula to benefit from both, setting it apart from the prior late materialization techniques for GPU query processing like the one in GHive~\cite{GHive}.
Hybrid systems can incorporate our optimization to improve their performance further.

\noindent
\textbf{GPU slicing for workload collocation.}
Some prior work~\cite{robroek2024euromlsys, cao2024vldb} explores GPU slicing techniques to collocate workloads that stress different micro-architecture resources on a single GPU.
Resources, like memory bandwidth, that would be left idle by a workload can be used by another complementary workload. 
\THISWORK's IO redistribution idea at the GPU level, can be combined with GPU slicing at sub-GPU granularity for even higher overall system efficiency, which we leave as future work.
For example, one can assign a 25\% GPU slice with all GPUs' IO bandwidth with \THISWORK's techniques to further resolve the IO bottleneck.

\section{Conclusion} \label{section:conclusion}
This paper introduced \THISWORK, a fundamental rethinking of how to overcome the limited memory capacity in GPUs for large-scale data analytics. 
A key component of \THISWORK~is an optimized IO primitive that utilizes all PCIe links in a multi-GPU system to transfer data to the GPU executing IO-intensive analytics. 
We presented a novel programming model that enables independent GPU code development and reuse, decoupled from IO scheduling and management.
We demonstrated data partitioning and compute scheduling for workloads that exceed GPU memory capacity through specific query operators, as well as a late materialization technique based on GPU zero-copy memory access that further improves query performance.
Using end-to-end SSB queries, we achieved a performance gain of 5.7$\times$ over state-of-the-art GPU baselines and an improved price-performance of 2.5$\times$ compared to a CPU baseline.

\begin{acks}
This work was supported in part by Advanced Micro Devices, Inc. under the ``Funding Academic Research" and the AMD AI \& HPC Cluster Program.
\end{acks}


\newpage

\onecolumn \begin{multicols}{2}
\bibliographystyle{ACM-Reference-Format}
\bibliography{00_main}
\end{multicols}

\end{document}